\documentclass[11pt,conference]{IEEEtran}

\usepackage{graphics}
\usepackage{amssymb}
\usepackage{amscd}
\usepackage{amsmath}
\usepackage{epsfig}
\usepackage{rotating}
\usepackage{amsmath}
\usepackage{amsfonts}
\usepackage{url}

\usepackage[english]{babel}
\renewcommand{\baselinestretch}{.93}
\begin{document}
\title{Opportunistic Interference Alignment in MIMO Interference Channels}
\author{\IEEEauthorblockN{Samir Medina Perlaza$^{1}$, M\'{e}rouane Debbah$^{2}$, Samson Lasaulce$^{3}$ and Jean-Marie Chaufray$^{1}$}
\IEEEauthorblockA{\\$^{1}$ France Telecom R\&D - Orange Labs Paris. France\\
$\{$Samir.MedinaPerlaza, JeanMarie.Chaufray$\}$@orange-ftgroup.com\\
$^{2}$ Alcatel Lucent Chair in Flexible Radio - SUPELEC. France\\
Merouane.Debbah@supelec.fr\\
$^{3}$ Laboratoire des Signaux et Syst\`emes (LSS) - CNRS, SUPELEC, Univ. Paris Sud.  France\\
Samson.Lasaulce@lss.supelec.fr}\\
}
\maketitle

\begin{abstract}
\boldmath We present two interference alignment techniques such that an opportunistic point-to-point multiple input multiple output
(MIMO) link can reuse, without generating any additional interference, the same frequency band of a similar pre-existing primary
link. In this scenario, we exploit the fact that under power constraints, although each radio maximizes independently its rate by water-filling on their channel transfer matrix singular values, frequently, not all of them are used. Therefore, by aligning the interference of the
opportunistic radio it is possible to transmit at a significant rate while insuring zero-interference on the pre-existing link. We
propose a linear pre-coder for a perfect interference alignment and a power allocation scheme which maximizes the individual data rate
of the secondary link. Our numerical results show that significant data rates are achieved even for a reduced number of antennas.
\end{abstract}

\section{Introduction}\label{SecIntroduction}
We consider the case of radio devices attempting to
opportunistically exploit the same frequency band being utilized by
licensed networks under the constraint that no additional
interference must be generated. This can be implemented for example
by assuming that opportunistic users are cognitive
radios \cite{Haykin2005, fette-book-2006}. Typically, cognitive
radios temporally
 exploit the unused frequency bands, named white-spaces, to
  transmit their data. This clearly improves the spectral
   efficiency since more users are allowed to co-exist in the
   same bandwidth. However, a higher spectrum
   efficiency could be attained by simultaneously allowing the
    opportunistic radios to transmit with the licensees
    if no harmful interference is generated. In this case,
     interference alignment (IA) has been identified as a
      powerful tool to achieve such a goal. This technique, was initially introduced in \cite{Khandani2006Report, Khandani2006}. IA allows a
       given transmitter to partially or completely  ``align''
       its interference with unused dimensions of the primary
       terminals \cite{Jafar2008Alignment}. The concept of dimension could be associated
       with a specific spatial direction, frequency carrier or
       time slot \cite{Khandani2006}, \cite{Jafar2007}. An
       extensive study has been conducted \cite{Khandani2006, Jafar2007, JafarXChannel2008} to
       estimate the number of interference-free dimensions a given radio might
       find to transmit when several radio systems co-exist. In 
       \cite{Khandani2006} and \cite{Jafar2007} several IA schemes to exploit such dimensions are proposed. Similarly, in \cite{Cardoso2008}, a linear pre-coder based on Vandermonde matrices allows an
       orthogonal frequency division multiplexing (OFDM) radio to
       co-exist with similar pre-existing terminals without generating any additional interference. The idea is to exploit the redundancy of the OFDM cyclic prefix and frequency selectivity of the channel.
\begin{figure}
\begin{center}
\includegraphics[width=.6\linewidth]{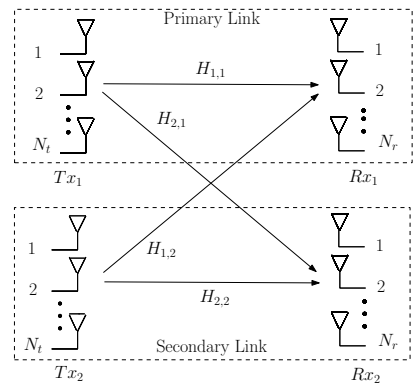}
\caption{\label{FigSystemModel} Two-user MIMO interference channel.}
\end{center}
\end{figure}

In this study, we propose a novel interference alignment technique
for the secondary users exploiting the fact that under a power-limitation, a primary user which maximizes its own rate
  by water-filling on its MIMO channel singular values, might leave some of them unused. \emph{i.e.} no transmission takes place along the corresponding spatial directions. These unused directions may be opportunistically utilized by a secondary transmitter, since its signal would not
  interfere with the signal sent by the primary transmitter. We present a linear
  pre-coder which perfectly aligns the interference generated by the
  secondary transmitter with such unused spatial directions. Similarly,
  we present a power allocation scheme based on the water-filling idea
  which maximizes the individual data rate of the opportunistic radio. Simulation
   results show that a significant data rate can be achieved by the secondary
   link following our approach.

\section{System Model}\label{SecSystemModel}
\emph{Notation:} In the following, matrices and vectors are denoted
by boldface upper case symbols and boldface lower case symbols, respectively. The
 $i^{th}$ entry of the vector $\boldsymbol{x}$ is denoted $\boldsymbol{x}(i)$. The
 entry corresponding to the $i^{th}$ row and $j^{th}$ column of the matrix $\boldsymbol{X}$ is
 denoted by  $\boldsymbol{X}(i,j)$. The $N$-dimension identity and null matrix are
 represented by $\boldsymbol{I}_N$ and $\boldsymbol{0}_N$, respectively. The
 Hermitian transpose is denoted $(\cdot)^H$, and the expected value is represented by the operator $\mathbb{E}\left[.\right]$.

We consider two point-to-point unidirectional links simultaneously
operating in the same frequency band and producing mutual interference as
shown in Fig. \ref{FigSystemModel} \cite{Sato1977}. Both transmitters are equipped
with $N_t$ antennas while both receivers use $N_r$ antennas. The first transmitter-receiver pair, i.e. $Tx_1$ and $Rx_1$, is a primary link licensed to
exclusively exploit a given frequency band. The pair, $Tx_2$ and $Rx_2$ is
an opportunistic link exploiting the same frequency band subject
to the constraint that no additional interference must be generated
over the primary system. Note that no cooperation between
terminals is allowed, \emph{i.e.} transmitters do not share or
exchange any signal before transmitting. Therefore, the multiple
access interference (MAI) is considered as additive white Gaussian
noise (AWGN).

The channel transfer matrix from transmitter $j  \in \left\lbrace 1,
2 \right\rbrace$ to receiver $i  \in \left\lbrace 1, 2
\right\rbrace$ is denoted  $\boldsymbol{H}_{i,j} \in
\mathbb{C}^{N_r\times N_t}$, where the entries of
$\boldsymbol{H}_{i,j}$  are independent and identically distributed
(i.i.d) complex Gaussian circularly symmetric random variables. The
channel matrices are supposed to be fixed for the whole transmission
duration. This correspond to assuming (static) Gaussian links. But
our analysis readily extends to the case of slow-fading channels by
assuming the channels to be constant over each data block. Regarding
the channel state information (CSI) conditions, we assume the
primary terminals (transmitter and receiver) to only have perfect
 knowledge of the matrix $\boldsymbol{H}_{1,1}$. On the other hand,
 the secondary terminals have perfect knowledge of all the channel
 transfer matrices $\boldsymbol{H}_{i,j}$, for every $i$
 and $j \in \left\lbrace 1, 2 \right\rbrace$. Although unrealistic,
 this condition provides us with an upper bound on the achievable
 rate of the secondary user. It can however be met in practice in the TDD (Time Division Duplex) mode if the secondary
 user exploits opportunistically the training sequences and signaling communication between the
 primary devices.

Following a matrix notation, the primary and secondary received
signals can be written as
\begin{equation}\label{EqReceivedSignal}
\left( \begin{array}{c} \boldsymbol{y}_1 \\
\boldsymbol{y}_2\end{array}\right) = \left( \begin{array}{cc}
\boldsymbol{H}_{1,1} & \boldsymbol{H}_{1,2} \\ \boldsymbol{H}_{2,1}
& \boldsymbol{H}_{2,2} \end{array}\right) \left(\begin{array}{c}
\boldsymbol{V}_1  \boldsymbol{s}_1 \\ \boldsymbol{V}_2
\boldsymbol{s}_2 \end{array}\right) + \left( \begin{array}{c}
\boldsymbol{n}_1 \\ \boldsymbol{n}_2 \end{array}\right),
\end{equation}
where the vectors $\boldsymbol{s}_i \in \mathbb{C}^{N_t \times 1}$ and
  $\boldsymbol{n}_i \in \mathbb{C}^{N_r \times 1}$ represent the transmitted
  symbols and an AWGN process with zero mean and covariance matrix $\sigma^2 \boldsymbol{I}_{N_r}$ for
  the $i^{th}$ link. For all $i \in \left\lbrace 1,2 \right\rbrace$ the
   matrices $\boldsymbol{V}_i \in \mathbb{C}^{N_t \times N_t}$ represent the linear pre-coders used
    for interference alignment. Furthermore, at each receiver, the
    input signal is linearly processed with the matrix
    $\boldsymbol{F}_i \in \mathbb{C}^{N_r \times N_r}$. The signal at the output of the linear filter $i$ is $\boldsymbol{r}_i = \boldsymbol{F}_i
    \boldsymbol{y}_i$. Both matrices $\boldsymbol{V}_i$ and $\boldsymbol{F}_i$ are
    described
    later on. The power allocation matrices are defined as the input covariance
matrices $\boldsymbol{P}_i = \mathbb{E}\left[\boldsymbol{s}_i
\boldsymbol{s}_i^H \right] \in \left(\mathbb{R}^+ \right)^{N_t
\times N_t}$, for the $i^{th}$ transmitter. The power constraints
are
\begin{equation}\label{EqPowerConstraints}
\forall i \in \left\lbrace 1, 2 \right\rbrace, \quad \text{Trace}\left( \boldsymbol{V}_i \boldsymbol{P}_{i} \boldsymbol{V}_i^H \right) \leqslant p_{i, \mathrm{max}},
\end{equation}
where $p_{i, \mathrm{max}}$ is the maximum transmit
 power level for the $i^{th}$ transmitter. Without loss of generality, we assume
 identical maximum transmit powers for
 all the terminals \emph{i.e.} $\forall i \in \left\lbrace 1,2 \right\rbrace$, $p_{i, \mathrm{max}} = p_{\mathrm{max}}$.

\section{Interference Alignment Strategy} \label{SecStrategy}

In this section we focus on the study of the pre-coding
$\boldsymbol{V}_i$ and post-processing $\boldsymbol{F}_i$ matrices.
Suppose that  the primary terminals completely ignore the presence
of the opportunistic transmitter. Hence, in order to maximize its
own data rate, the primary transmitter follows a water-filling power
allocation as in the single-user case \cite{Cioffi2004}.
\subsection{Primary link design}
Under the assumption that the channel matrix
$\boldsymbol{H}_{1,1}$ is known at the receiver and transmitter, the
primary terminal chooses its pre-coding $\boldsymbol{V}_1$ and
post-processing $\boldsymbol{F}_1$ matrices in such a way that their
channel transfer matrix is diagonalized, \emph{i.e.}
$\boldsymbol{V}_1$ and $\boldsymbol{F}_1 = \boldsymbol{U}_1^H$
satisfy the singular value decomposition $\boldsymbol{H}_{1,1} =
\boldsymbol{U}_1 \boldsymbol{\Lambda}_1 \boldsymbol{V}^{H}_1$, where
$\boldsymbol{U}_1 \in \mathbb{C}^{N_r \times N_r}$ and
$\boldsymbol{V}_1 \in \mathbb{C}^{N_t \times N_t}$ are unitary
matrices and $\boldsymbol{\Lambda_1} \in
\left(\mathbb{R}^+\right)^{N_r \times N_t}$ is a diagonal matrix
which contains $\mathrm{min}(N_r,N_t)$ non-zero singular values, $
\lambda_1,\ldots,\lambda_{\mathrm{min}(N_r,N_t)}$. Thus, the received signal after
linear processing, $\boldsymbol{r}_1$, can be written as
\begin{equation}\label{EqWhitenedReceivedSignal}
\boldsymbol{r}_1 = \boldsymbol{F}_1 \boldsymbol{y}_1 = \boldsymbol{\Lambda}_1 \boldsymbol{s}_1  + \boldsymbol{n}_1',
\end{equation}
where $\boldsymbol{n}_1' = \boldsymbol{U}^H \boldsymbol{n}_1$ is an
AWGN process with zero mean and covariance matrix $\sigma^2 {\bf
I}_{N_r}$. Then, the achievable rate of the primary user is
maximized by the power allocation matrix $\boldsymbol{P}_1$ which is
a solution to the following optimization problem
\begin{equation}\label{EqOptimizationPrimaryUser}
\begin{array}{lc}
\text{maximize} &   \log_2 \left|\boldsymbol{I}_{N_t} + \frac{1}{\sigma^2} \boldsymbol{H}_{1,1}  \boldsymbol{V}_1 \boldsymbol{P}_1 \boldsymbol{V}_1^H \boldsymbol{H}_{1,1}^H  \right|\\
\text{subject to} & \text{Trace}\left( \boldsymbol{P}_1 \right) \leqslant p_{\mathrm{max}}.
\end{array}
\end{equation}
The solution to (\ref{EqOptimizationPrimaryUser}) is the classical water-filling algorithm
\cite{Cioffi2004}. Following this approach, the optimal power
allocation matrix is a diagonal matrix with entries
\begin{equation}\label{Eqwater-filling}
\forall n \in \left\lbrace 1,\ldots,N_t \right\rbrace, \quad
P_1(n,n) = \left[\beta - \frac{\sigma^2}{\lambda_n^2}\right]^+,
\end{equation}
with $\left[ p \right]^+ = \max\left(0,p\right)$. The constant $\beta$ is a Lagrangian multiplier that is determined to satisfy
\begin{equation*}
\displaystyle\sum_{j =1}^{N} \boldsymbol{P}_1(j,j) = p_{\mathrm{max}}.
\end{equation*}
\subsection{Secondary link design}
Depending on the channel singular values $\lambda_1, \ldots,
\lambda_{\mathrm{min}(N_r,N_t)}$, the power allocation matrix $\boldsymbol{P}_1$ might
contain  zeros in its main diagonal. A zero power allocation for a
given singular value means that no transmission takes place along
the corresponding spatial direction. This means that the secondary
terminal can align its transmitted signal with the unused singular
modes such that it does not interfere with the signal transmitted by
the primary user. If one converts the spatial problem into the
frequency one, the result is similar to the cognitive scenario where
the secondary would opportunistically use the unexploited frequency
modes. The main difference here lies in the fact that in the spatial
domain, there is no universal precoder which diagonalizes the basis
of all the devices whereas this is the case in the frequency domain
with the use of the FFT.

As a consequence, in the spatial domain, the corresponding
orthogonality condition (such that the secondary user generates no
interference on the primary link) is given by
\begin{equation}\label{EqOrthogonalityCondition}
\boldsymbol{U}_1^H \boldsymbol{H}_{1,2} \boldsymbol{V}_{2} = \alpha
\bar{\boldsymbol{P}_1},
\end{equation}
where the matrix $\bar{\boldsymbol{P}_1}$ is a diagonal matrix with entries
\begin{equation}\label{EqComplementaryPowerAllocation}
 \forall n \in \left\lbrace 1,\ldots,N_t\right\rbrace, \quad \boldsymbol{\bar{P_1}}(n,n) = \left[\frac{\sigma^2}{\lambda_n^2} - \beta \right]^+,
\end{equation}
such that the condition $\boldsymbol{P}_1 \bar{\boldsymbol{P}}_1 =
\boldsymbol{0}_{N_r}$ always holds. It can be easily verified since both matrices are diagonal.
Additionally, the constant $\alpha$ is chosen
to satisfy the power constraints (\ref{EqPowerConstraints}) with $i
= 2$.

Assuming that perfect estimates of $\boldsymbol{H}_{1,1}$ and
$\boldsymbol{H}_{1,2}$ are available at the secondary transmitter,
the secondary precoder (when the inverse of $\boldsymbol{H}_{1,2}$
exists) is given by:
\begin{equation}\label{EqPrecoder}
\boldsymbol{V}_2 = \alpha \boldsymbol{H}_{1,2}^{-1}
\boldsymbol{U}_{1} \bar{\boldsymbol{P}}_1.
\end{equation}
For the case where $N_r > N_t$, \emph{i.e.} the receiver has more
antennas than the transmitter, it is still possible to obtain the
pre-coding matrix by using the Moore-Penrose pseudo-inverse of
$\boldsymbol{H}_{1,2}$,
\begin{equation}\label{EqPrecoderIllConditioned}
\boldsymbol{V}_2 =  \alpha \left( \boldsymbol{H}_{1,2}^H
\boldsymbol{H}_{1,2} \right)^{-1}\boldsymbol{H}_{1,2}^H
\boldsymbol{U}_{1}  \bar{\boldsymbol{P}}_1.
\end{equation}

Once the pre-decoder $\boldsymbol{V}_2$ has been adapted to satisfy
(\ref{EqPrecoder}) or (\ref{EqPrecoderIllConditioned}) at the
secondary transmitter, no additional interference impairs the
primary user. However, the secondary receiver still undergoes the
interference from the primary transmitter. Typically, this effect is
a colored noise with covariance $\boldsymbol{Q} \in \mathbb{C}^{N_r
\times N_r}$ due to the channel $\boldsymbol{H}_{2,1}$ and the
pre-coder $\boldsymbol{V}_1$. Here,
\begin{equation}\label{EqMAINCovariance}
\boldsymbol{Q} = \boldsymbol{H}_{2,1} \boldsymbol{V}_1
\boldsymbol{P}_1 \boldsymbol{V}_1^H \boldsymbol{H}_{2,1}^H  +
\sigma^2 \boldsymbol{I}_{N_r}.
\end{equation}
Hence, the received signal $\boldsymbol{y}_2$ can be whitened  by
using the
 matrix $\boldsymbol{F}_2 = \boldsymbol{Q}^{-\frac{1}{2}}$, to
 obtain $\boldsymbol{r}_2 =\boldsymbol{F}_2 \boldsymbol{y}_2$, such that
\begin{equation}\label{EqFilteredSignal2}
\boldsymbol{r}_2 = \boldsymbol{Q}^{-\frac{1}{2}}
\boldsymbol{H}_{2,2} \boldsymbol{V}_2  \boldsymbol{s}_2 +
\boldsymbol{n}_2',
\end{equation}
where $\boldsymbol{n}_2' = \boldsymbol{Q}^{-\frac{1}{2}} \left(
\boldsymbol{H}_{2,1}  \boldsymbol{V}_1 \boldsymbol{s}_1 +
\boldsymbol{n}_2 \right)$ is an i.i.d. AWGN process with zero mean
and a covariance matrix proportional to the identity. Let $S$ be the
number of zeros on the main diagonal of $\boldsymbol{P}_1$. Then,
the matrix $\boldsymbol{V}_2$ contains $N_{r} - S$ zero columns.
Note that $S = 0$ implies that no transmission takes
  place in the secondary link. In the sequel, we always assume that $S > 0$ (which will be the case at low signal to noise ratio as shown in the simulations).

\section{Input Covariance Matrix Optimization}\label{SecInputCovarianceMatrix}
In the latter section the proposed pre-coding scheme does not generate
any interference on the primary user but the transmission rate for
the secondary user was not optimized. For this purpose, the choice
of the power allocation of the secondary transmitter, \emph{i.e.} the matrix ${\bf P}_2$, needs to be optimized. First, we present the most
simple case where uniform power allocation is performed. Second, we
introduce a power allocation which maximizes the individual
transmission rate. In both cases we assume that the pre-coder has
been previously adapted to satisfy the orthogonality conditions
(\ref{EqPrecoder}) or (\ref{EqPrecoderIllConditioned}).

\subsection{Uniform Power Allocation}

For the uniform power allocation scheme the input covariance matrix
is set to $\boldsymbol{P}_2 = \boldsymbol{I}_{N_t}$ and the constant
$\alpha$ from (\ref{EqOrthogonalityCondition}) is tuned in order to
meet the condition $\text{Trace}\left(\boldsymbol{V}_2
\boldsymbol{V}_2^H \right) = p_{\mathrm{max}}$. The rate achieved by the
secondary user while generating zero-interference to the primary
receiver is
\begin{equation}\label{EqSecondaryRate}
R_2 = \log_2 \left| \boldsymbol{I}_{N_r} + \boldsymbol{Q}^{-\frac{1}{2}} \boldsymbol{H}_{2,2} \boldsymbol{V}_2  \boldsymbol{V}_2^H  \boldsymbol{H}_{2,2}^H \boldsymbol{Q}^{-\frac{1}{2}} \right|.
\end{equation}

\subsection{Optimal Power Allocation}
The transmission rate for the secondary link is maximized by
adopting a power allocation matrix $\boldsymbol{P}_2$ which is a
solution of the following optimization problem,
\begin{equation}\label{EqOptimizationProblem2}
\begin{array}{lc}
\displaystyle\arg\max_{\boldsymbol{P}_2} & R_2(\boldsymbol{P}_2)\\
\text{s.t.} & \text{Trace}\left( \boldsymbol{V}_2 \boldsymbol{P}_2
 \boldsymbol{V}_2^H \right) \leqslant p_{\mathrm{max}},
\end{array}
\end{equation}
where
\begin{equation}
R_2(\boldsymbol{P}_2) = \log_2 \left|\boldsymbol{I}_N +
\boldsymbol{Q}^{-\frac{1}{2}} \boldsymbol{H}_{2,2}
 \boldsymbol{V}_2 \boldsymbol{P}_2 \boldsymbol{V}_2^H
 \boldsymbol{H}_{2,2}^H \boldsymbol{Q}^{-\frac{1}{2}}  \right|.
\end{equation}
Note that
  solving this optimization problem requires the knowledge of
  the covariance matrix $\boldsymbol{Q}$, which is calculated at
  the secondary receiver based on the knowledge of the
  channel $\boldsymbol{H}_{2,1}$. This can be done if  the
  secondary receiver estimates $\boldsymbol{Q}$ and feeds
  it back to the secondary transmitter. Here, we assume a perfect
      knowledge of $\boldsymbol{Q}$ is available at the secondary transmitter.

By definition (Eq. (\ref{EqOptimizationProblem2})), the matrix $
\boldsymbol{V}_2$ is not full rank. Therefore, the optimization
problem (\ref{EqOptimizationProblem2}) does not have a simple
solution. We propose a two-step optimization which leads to a
water-filling solution. First, we define a new input covariance
$\hat{\boldsymbol{P}}_2$ such that,
\begin{equation}\label{EqNewInputCovariance}
\hat{\boldsymbol{P}_2} =  \left(\boldsymbol{V}_2^H
 \boldsymbol{V}_2 \right)^{\frac{1}{2}} \boldsymbol{P}_2
  \left(\boldsymbol{V}_2^H \boldsymbol{V}_2 \right)^{\frac{1}{2}}.
\end{equation}
By replacing the expression (\ref{EqNewInputCovariance}) in
 (\ref{EqOptimizationProblem2}), the optimization problem becomes
\begin{equation}\label{}
\begin{array}{lc}
\displaystyle\arg\max_{\hat{\boldsymbol{P}}_2}  & \left\lbrace
 \log_2 \left|\boldsymbol{I}_{N_r} + \boldsymbol{G} \hat{\boldsymbol{P}}_2
 \boldsymbol{G}^H  \right| \right\rbrace \\
\text{s.t} & \text{Trace}\left( \hat{\boldsymbol{P}}_2 \right) = p_{\mathrm{max}},
\end{array}
\end{equation}
where $\boldsymbol{G} = \boldsymbol{Q}^{-\frac{1}{2}} \boldsymbol{H}_{2,2}
 \boldsymbol{V}_{2} \left( \boldsymbol{V}_2^H  \boldsymbol{V}_2 \right)^{-\frac{1}{2}}
 \in \mathbb{C}^{N_r \times N_t}$. The idea here is to solve the a
 priori non-trivial optimization problem defined by expression (\ref{EqOptimizationProblem2}) by
 introducing an equivalent channel matrix $\boldsymbol{G}$ to
 simplify the problem. Using $\boldsymbol{G}$ we can then apply a singular value decomposition to the new channel such that
$\boldsymbol{G} = \boldsymbol{E} \boldsymbol{\Delta}
\boldsymbol{Z}^H$, where $\boldsymbol{E} \in \mathbb{C}^{N_r \times
N_r}$ and $\boldsymbol{Z} \in \mathbb{C}^{N_t \times N_t}$ are unitary
matrices, and the matrix $\boldsymbol{\Delta} \in \left(\mathbb{R^+}\right)^{N_r
\times N_t}$ contains the singular values $\eta_1, \dots, \eta_{\min\left\lbrace N_r, N_t \right\rbrace}$ of
$\boldsymbol{G}$. Under these assumptions, the optimal solution
$\boldsymbol{P}_2^* = \boldsymbol{Z}^H \hat{\boldsymbol{P}}_2
\boldsymbol{Z}$ is
\begin{equation}\label{Eqwater-filling2}
\forall n \in \left\lbrace 1,\ldots,N_t\right\rbrace, \quad \boldsymbol{P}_2^*(n,n) = \left[\rho - \frac{1}{\eta_n^2}\right]^+,
\end{equation}
where $\rho$ is a Lagrangian multiplier that is determined to
satisfy $\displaystyle\sum_{j =1}^{N} \boldsymbol{P}_2^*(j,j) = p_{\mathrm{max}}$. Once $\boldsymbol{P}_2^*$ has been obtained, then the optimal power allocation matrix \cite{Cioffi2004} is
\begin{equation}\label{EqOptimalPower2}
\boldsymbol{P}_2 = \left(\boldsymbol{V}_2^H \boldsymbol{V}_2 \right)^{-\frac{1}{2}}
 \boldsymbol{Z} \hat{\boldsymbol{P}}_2^* \boldsymbol{Z}^H
  \left(\boldsymbol{V}_2^H \boldsymbol{V}_2  \right)^{-\frac{1}{2}}
\end{equation}
The constant $\alpha$ in (\ref{EqOrthogonalityCondition}) is tuned
such that the condition (\ref{EqPowerConstraints}) is met for $i =
2$.

\section{Numerical Results}
In this section we show numerical examples to illustrate
 the performance of our interference alignment strategy.
Considering the same number of antennas at the receiver and
 the transmitter, we analyze the number of unused singular
 values or free dimensions available for the secondary link
  as well as its achieved data rate. Recall that the primary
  link is interference-free, therefore its data rate corresponds
  to the single user case rate \cite{Cioffi2004}.

\begin{figure}
\begin{center}
\includegraphics[width=\linewidth]{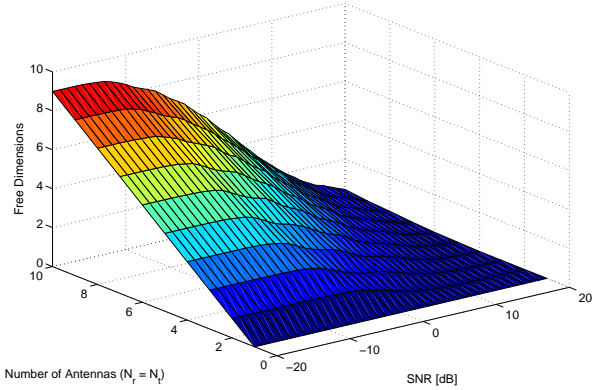}
\caption{\label{FigSingularValues} Average number of unused
singular values in the primary link as a function of the total number of antennas and the $SNR = \frac{p_{\mathrm{max}}}{\sigma^2}$. The SNR and the number of antennas $N_r = N_t$ are assumed the same for the primary and secondary links}
\end{center}
\end{figure}

\begin{figure}
\begin{center}
\includegraphics[width=\linewidth]{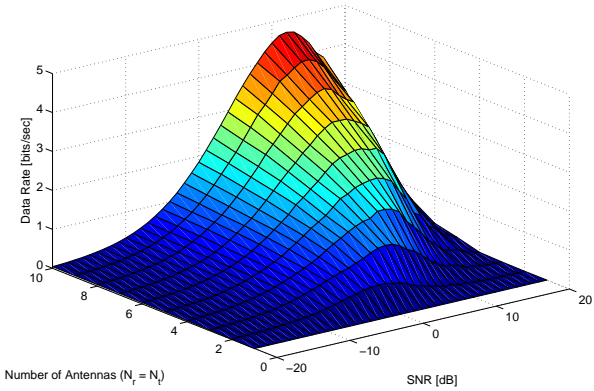}
\caption{\label{FigDataRateOptimal} Average data rate of the secondary link when optimal power allocation is implemented as a function of the number of antennas $N_r = N_t$ and $SNR = \frac{p_{\mathrm{max}}}{\sigma^2}$. The SNR and the number of antennas are assumed the same for the primary and secondary link.}
\end{center}
\end{figure}

\begin{figure}
\begin{center}
\includegraphics[width=\linewidth]{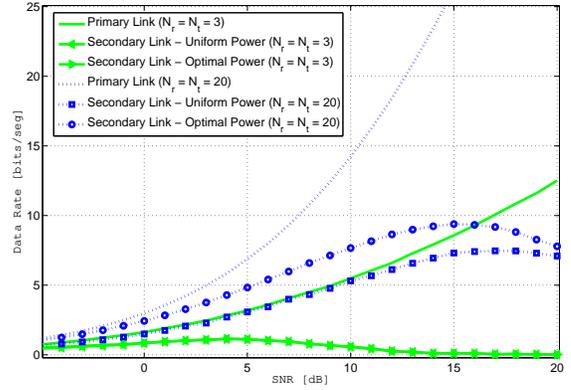}
\caption{\label{FigDiffR12Optimal} Average data rate achieved by the primary $R_{1}$ and secondary link for uniform $R_{\mathrm{2,uniform}}$ and optimal $R_{\mathrm{2,optimal}}$ power allocation as a function of their $SNR = \frac{p_{\mathrm{max}}}{\sigma^2}$. The dashed lines correspond to $N_r = N_t = 20$ antennas. The solid lines correspond to $N_r = N_t = 3$ antennas.}
\end{center}
\end{figure}

In Fig. \ref{FigSingularValues} we show the number of unused
singular values in the primary link as a function of the number of
antennas and $SNR = \frac{p_{\mathrm{max}}}{\sigma^2}$. Note that in
low SNR regime, the transmitter attempts to concentrate all its
power in the best singular values leaving all the others unused. On
the contrary, in high SNR regime the primary transmitter tends to
spread its power among all its available singular values. Thus, in
the first case the opportunistic link has plenty of free dimensions,
while in the second one, it is effectively limited. This power
allocation behavior has been also reported in \cite{Valenzuela2002}
and \cite{Valenzuela2002a}. Similarly, it is observed that
increasing the number of antennas leads on average, to a linear
scaling of the unused singular values. In Fig.
\ref{FigDataRateOptimal}, we show the achieved data rate of
secondary link when optimal power allocation is implemented
($R_{\mathrm{2,optimal}}$) as a function of the number of antennas
and the SNR. Therein, it is shown that at very low and very high
SNRs the data rate approaches zero bits/sec. In low SNR regime this
effect is natural since detection is difficult due to the noise.
However, at high SNRs it is due to the fact that the primary
transmitter does not leave any unused singular value. Nonetheless,
at intermediate SNRs, significant data rates are achieved by the
secondary link. Note that increasing the number of antennas always
leads to higher data rates for the secondary link. In Fig.
\ref{FigDiffR12Optimal}, we plot the data rate of the primary $R_{\mathrm{1}}$ and secondary link
for the case of uniform $R_{\mathrm{2,uniform}}$ and optimal $R_{\mathrm{2,optimal}}$ power allocation. Note that in high
SNR regime, the $R_{\mathrm{2,uniform}}$ and $R_{\mathrm{2,optimal}}$
performs similarly. It is due to the fact that few or even none of
the singular values are left unused by the primary link, therefore the uniform power allocation does not differ
from the optimal in average. In contrast, at intermediate SNRs the
difference in performance is more significant for a large number of antennas. 

\section{Conclusions}\label{SecConclusions}

We provided a novel interference alignment scheme which allows an opportunistic
point-to-point MIMO link to co-exist with a similar pre-existing primary link  on the same
fully utilized band without generating any additional interference. The proposed scheme
exploits the fact that the licensed transmitter, while performing  water-filling power
allocation on its MIMO channel (in order to maximize its single user rate)  will leave
some singular values unused when constrained by power limitations.  Hence, no
transmission takes place along the corresponding spatial directions. We proposed a linear
pre-coder for the opportunistic radio, which perfectly aligns the transmitted signal with
such unused dimensions. We also provided a power allocation scheme which maximizes the
data rate of the opportunistic link. Numerical results show that significant data rates
are achieved by the secondary link even for a reduced number of antennas. Further studies
will extend the novel  approach to multi-user multi-carrier systems and  the case of
incomplete CSI.
\bibliographystyle{IEEEtran}
\bibliography{spectrumsharing}

\end{document}